\documentclass[
twocolumn,
superscriptaddress, showpacs, nofootinbib
, longbibliography
]{revtex4-1}

\usepackage{graphicx}
\usepackage{amsmath,amssymb,amsfonts}
\usepackage{bm}
\usepackage{color}
\usepackage{comment}
\usepackage{slashed} 
\usepackage{lineno}

\allowdisplaybreaks[3]

\newcommand{\nn}{\nonumber}

\newcommand{\gam}{ \gamma }

\newcommand{\ep}{ \epsilon }
\newcommand{\vep}{ \varepsilon }
\newcommand{\Lam}{ \Lambda }

\newcommand{\Lag}{ {\mathcal L} }
\newcommand{\M}{ {\mathcal M} }

\newcommand{\T}{ {\mathcal T}  }
\newcommand{\prj}{ {\mathcal P} }
\newcommand{\Q}{ {\mathcal Q} }

\newcommand{\bx}{{\bm{x}}}

\newcommand{\bk}{{\bm{k}}}
\newcommand{\bp}{{\bm{p}}}

\newcommand{\bv}{{\bm{v}}}
\newcommand{\bl}{{\bm{\ell}}}
\newcommand{\bA}{{\bm{A}}}
\newcommand{\bgam}{ {\bm \gamma} }

\newcommand{\sla}{\slashed} 
\newcommand{\para}{\parallel}

\newcommand{\SC}{{\rm 2SC}}
\newcommand{\U}{ {\mathcal U} }
\newcommand{\order}{ {\mathcal O} }

\usepackage[normalem]{ulem}  

\renewcommand\sout{\bgroup \color{red} \ULdepth=-.5ex \ULset}

\usepackage[
breaklinks, bookmarks, pdfstartview=FitV, 
colorlinks, linkcolor=blue, urlcolor=cyan,citecolor=red]{hyperref}

\begin{document}

\begin{flushright}
YITP-19-21
\end{flushright}

\title{
%
%
Emergent QCD Kondo effect in two-flavor color superconducting phase
}

\author{Koichi Hattori}
\affiliation{Yukawa Institute for Theoretical Physics, Kyoto University, Kyoto 606-8502, Japan}
\affiliation{Physics Department and Center for Particle Physics and Field Theory, Fudan University, Shanghai 200433, China}

\author{Xu-Guang Huang}
\affiliation{Physics Department and Center for Particle Physics and Field Theory, Fudan University, Shanghai 200433, China}
\affiliation{Key Laboratory of Nuclear Physics and Ion-beam Application (MOE), Fudan University,
Shanghai 200433, China}

\author{Robert D. Pisarski}
 \affiliation{Physics Department, Brookhaven National Laboratory, Upton,
 New York 11973-5000, U.S.A}

\date{ \today}

\begin{abstract}
We show that effective coupling strengths between ungapped and gapped quarks in 
the two-flavor color superconducting (2SC) phase are renormalized by logarithmic quantum corrections. 
We obtain a set of coupled renormalization-group (RG) equations 
for two distinct effective coupling strengths arising from gluon exchanges carrying different color charges. 
The diagram of RG flow suggests that both of the coupling strengths 
evolve into a strong-coupling regime as we decrease the energy scale toward the Fermi surface. 
This is a characteristic behavior observed in the Kondo effect, which has been known to occur 
in the presence of impurity scatterings via non-Abelian interactions. 
We propose a novel Kondo effect emerging without doped impurities, 
but with the gapped quasiexcitations and the residual SU(2) color subgroup intrinsic in the 2SC phase, 
which we call the ``{\it 2SC Kondo effect}.''

\end{abstract}

\maketitle


\section{Introduction}

The Kondo effect gives rise to rich physics from an emergent strong-coupling regime in the low-energy dynamics. 
In 1964, Jun Kondo pointed out that 
a long standing problem about an anomalous behavior in the resistivity of alloy 
originates from the quantum corrections to the impurity-scattering amplitudes~\cite{kondo1964resistance}. 
Namely, the coupling strength between conduction electrons and an impurity 
is enhanced due to contributions of the next-to-leading order scattering processes. 
The modification of the coupling strength is best captured by the concept of the renormalization group (RG), 
which inspired subsequent developments in this important concept~\cite{anderson1970poor, Anderson:1970, Wilson:1974mb}. 
The scaling argument clearly indicates that the existence of a Fermi surface is crucial 
for the Kondo effect to occur \cite{Polchinski:1992ed, HIO} 
as we will explain in a brief review part in the next section.

In recent years, the Kondo effect was applied to nuclear physics \cite{Yasui:2013xr, Hattori:2015hka}. 
Especially, possible realization of the {\it QCD Kondo effect} in dense quark matter was discussed 
when heavy-quark impurities are embedded in light-quark matter \cite{Hattori:2015hka}. 
The results of the perturbative RG analyses indicate that the effective interaction strength between 
light and heavy quarks evolves into a Landau pole despite the small value of QCD coupling constant at high density. 
Subsequent studies investigated further consequences of the QCD Kondo effect, 
including interplay/competition with color superconductivity \cite{Kanazawa:2016ihl}, 
formation of ``Kondo condensates'' and modification of QCD phase diagram \cite{Yasui:2016svc, Yasui:2016yet, Suzuki:2017gde, Yasui:2017izi}, 
nonperturbative aspects near the IR fixed point by conformal field theory \cite{Kimura:2016zyv, Kimura:2018vxj}, 
estimates of transport coefficients \cite{Yasui:2017bey}, 
and QCD equation of state for an application to neutron/quark star physics \cite{Macias:2019vbl}. 
It is also remarkable that light quarks have the same scaling dimensions 
in a dense system and in a strong magnetic field as a consequence of analogous effective dimensional reductions \cite{HIO}, 
and that a strong magnetic field alone induces the QCD Kondo effect even at zero density \cite{Ozaki:2015sya}.

In this paper, we propose a realization of the Kondo effect without (doped) impurities in QCD. 
We consider the RG evolution of effective coupling strengths between gapped and ungapped quarks 
appearing in the two-flavor color superconducting (2SC) phase \cite{Rapp:1997zu, Alford:1997zt}. 
In the 2SC phase with three colors, two of three color states of quarks are involved in the Cooper pairing 
and acquire a gap above the Fermi surface (cf., Fig.~\ref{fig:dispersions}), while the other color state 
remains gapless with a finite density of states at the Fermi surface (see, e.g., Refs.~\cite{Shovkovy:2004me, Alford:2007xm} for review articles). 
Therefore, the gapped quarks may play a role of impurities in the low-energy dynamics below the gap size, 
and the 2SC phase intrinsically has the necessary setup of the Kondo effect.

The formation of the diquark condensates breaks the SU(3) color symmetry group down to SU(2). 
Gluons belonging to the unbroken SU(2) subgroup are coupled only to the gapped quarks, 
so that they are decoupled from all quarks in the aforementioned low-energy regime, 
realizing a pure gluodynamics \cite{Rischke:2000cn}. 
Here, we can focus on the remaining five gluons which mediate 
the interactions between ungapped and gapped quarks. 
Those gluons have different properties depending on if they are associated with 
the diagonal or off-diagonal Gell-Mann matrices, 
so that we introduce two distinct effective coupling strengths.

We will derive coupled RG equations for these two effective coupling strengths 
from the next-to-leading order scattering amplitudes (cf., Fig.~\ref{fig:diagrams_2SC}), 
and obtain an RG-flow diagram in Fig.~\ref{fig:RG_flow}. 
The fate of the RG flow depends on the initial conditions for the RG equations. 
We take the tree-level coupling strengths as initial conditions 
which depend on the magnitudes of the Debye and Meissner masses 
in the 2SC phase \cite{Rischke:2000qz, Casalbuoni:2001ha}. 
Plugging the initial conditions evaluated with those quantities, 
we find that both of the coupling strengths evolve into strong-coupling regimes, 
but have different signs for attraction and repulsion.

This paper is organized as follows. 
In the next section, we provide a brief review of the QCD Kondo effect 
which will be useful to identify the basic mechanism of the Kondo effect. 
Then, we investigate a novel Kondo effect occurring in the 2SC phase in Sec.~\ref{sec:2SC}. 
We conclude the paper in Sec.~\ref{sec:summary} with discussions. 
Some useful properties of the high-density and heavy-quark effective field theories are 
briefly summarized in Appendix~\ref{sec:EFT}.

\section{Brief review of QCD Kondo effect}

\label{sec:review}

We first provide a brief review of the QCD Kondo effect \cite{Hattori:2015hka}, 
highlighting the essential points of the discussions given in a review article \cite{HIO}. 
These preliminary discussions will be useful to identify the necessary ingredients of the Kondo effect.

\subsection{High density effective field theory}

The Kondo effect is induced by low-energy excitations near the Fermi surface scattering off dilute impurities. 
Therefore, we use the high density effective field theory (HD-EFT) 
to extract such low-energy degrees of freedom from the full theory \cite{Hong:1998tn, Hong:1999ru, 
Casalbuoni:2000na, Beane:2000ms, Schafer:2003jn}.

We decompose the fermion momentum into a sum of the large Fermi momentum 
and the small residual momentum $ \ell^\mu = (\ell^0 , \bl)$ as 
\begin{eqnarray}
\label{eq:mom_decomp}
p^0 = \ell^0
\, , \quad 
\bp = \mu \bv_{F} + \bl
\, ,
\end{eqnarray}
where $ \mu $ and $\bv_F  $ are the chemical potential and the Fermi velocity, respectively, 
and $  \ell^0,|\ell| \ll \mu $.  
The energy $ p^0$ is measured from the Fermi surface. 
Then, at the leading order (LO) of $ 1/\mu $ expansion, 
the low-energy degrees of freedom are extracted from the Dirac Lagrangian as (see Appendix~\ref{sec:HDEFT}) 
\begin{eqnarray}
\Lag &=& \bar \psi(x) (i\sla D + \mu \gam^0) \psi(x)
\simeq \sum_{\bv_F} \bar \psi_+   i v_{F+}^\mu D_\mu \gam^0  \psi_+ 
\label{eq:ph}
\, ,
\end{eqnarray}
where $ v_{F\pm}^\mu \equiv (1, \pm \bv_F)$. 
We introduced the low-energy field $ \psi_+(\ell;\bv_F) \equiv \prj_+  \psi (\ell ; \bv_F) $ 
for particle and hole excitations around the Fermi momentum $ \mu \bv_F $, 
by the use of projection operators 
$\prj_\pm \equiv \frac{1}{2} ( 1 \pm \gam^0 \bv_F \cdot \bgam)$. 

\begin{figure}
     \begin{center}
              \includegraphics[width=0.6\hsize]{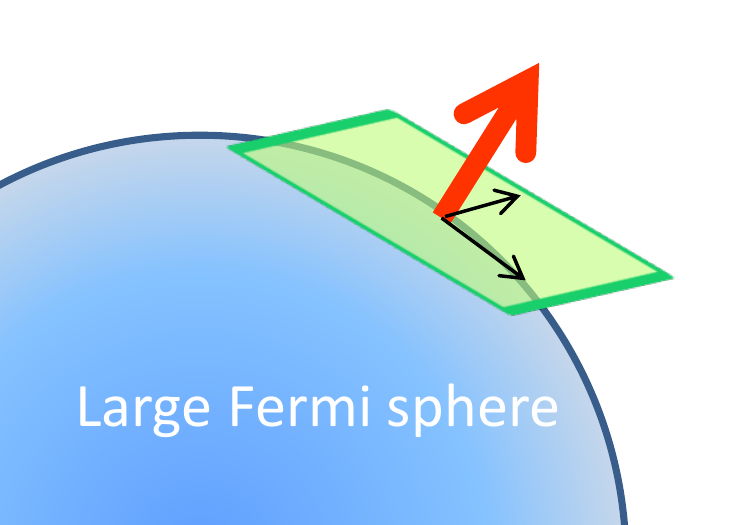}
     \end{center}
\vspace{-0.5cm}
\caption{Effective dimensional reduction near a large Fermi surface. 
Excitations cost energy only in the normal direction, 
so that there is a two-dimensional degeneracy in the tangential directions. 
}
\label{fig:FermiSurface}
\end{figure}

From this expansion, the dispersion relation of the low-energy excitations 
near the Fermi surface is read off as 
\begin{eqnarray}
 \ell^0 = \bv_F \cdot \bl 
 \, .
 \end{eqnarray} 
We define $ \ell_\para \equiv \bv_F \cdot \bl $ for later use. 
This is a linear dispersion relation in the (1+1)-dimensional phase space normal to the Fermi surface, 
and does not depend on the residual two-dimensional momentum tangential to the Fermi surface (cf., Fig.~\ref{fig:FermiSurface}). 
This means that an effective dimensional reduction occurs in the low-energy excitation near the large Fermi sphere, 
and the phase space is degenerated in the residual two dimensions. 

From Eq.~(\ref{eq:ph}), the free propagator is found to be 
\begin{eqnarray}
\label{eq:S_HD}
S (\ell; \bv_F) 
= \frac{ i \sla v_{F+}}{ 2 v_{F+}^{\mu} \ell_\mu + i \ell^0 \vep} 
= \frac{ i \, \prj_+ \gam^0 }{ v_{F+}^{\mu} \ell_\mu + i \ell^0 \vep} 
\, ,
\end{eqnarray}
where we have $ \sla v_{F\pm}  = 2 \prj_\pm \gam^0 $. 
At the LO, only the temporal and the parallel components of the gauge field, $A^0 $ and $ \bv_F \cdot \bA$, 
are coupled to the low-energy fermion excitations. 
The gamma matrix is not involved in these couplings, 
because the spin direction is frozen along the Fermi velocity. 

\subsection{Scaling dimensions in QCD Kondo effect}

One can determine the scaling dimension of the low-energy excitation field $\psi_+$ 
assuming that the kinetic term (\ref{eq:ph}) is invariant under the scaling transformation, 
$\ell^{0} \to s \ell^{0}$ ($t \to s^{-1}t$) with $  s<1$. 
According to the (1+1)-dimensional dispersion relation, 
only the $ \ell_\para $ scales as $\ell_\para \to s \ell_\para$, 
and the tangential momentum $ \bl_\perp  $ is intact. 
Therefore, the $\psi_+$ scales with a factor of $s^{-1/2}$ 
when the energy scale is reduced toward the Fermi energy ($  \ell^0 =0$).\footnote{
Since there is a degeneracy in the momentum space, 
it is useful to count the scaling dimensions in a mixed representation $ \psi_+(t, {\bm \ell}) $.}

In addition, we introduce a heavy-quark impurity embedded in the dense light-quark matter. 
One may use the heavy-quark effective theory (HQ-EFT) 
which is organized with an expansion with respect to the inverse heavy-quark mass $ 1/m_H $. 
This expansion is analogous to that in the HD-EFT as we summarize in Appendix~\ref{sec:HQEFT}. 
At the LO, the kinetic term of the particle state $\Psi_{+}$ is given by 
\begin{eqnarray}
S_{H}^{\rm{kin}}
= \int  dt \int \frac{ d^{3} \bk}{(2\pi)^3} \, \Psi_{+}^\dagger  \, i \partial_{t} \Psi_{+} + {\mathcal O}(1/m_H)
\label{kin_Q}
.
\end{eqnarray}
We consider a static heavy quark with a vanishing spatial velocity. 
In this case, the spatial derivative is not contained in the kinetic term, 
and we find that the heavy-quark field $\Psi_{+}$ and its spatial momentum $  \bk$ 
do not scale when $t \to s^{-1}t$. 
This is reasonable since the static impurity acts as a scattering center 
in the same way at any energy scale of light particles.

Now that we have determined the scaling dimensions of both the light and heavy quark fields, 
we look into the four-Fermi operator composed of the light and heavy quarks: 
\begin{eqnarray}
S^{\rm{int}}_{\rm HD}
&=& \int dt 
\sum_{\bv_F^{(1)},\bv_F^{(3)}} 
\int \frac{d^2\bl_\perp^{(1)} d\ell_\para^{(1)}} {(2\pi)^3} \frac{d^2\bl_\perp^{(3)} d\ell_\para^{(3)}} {(2\pi)^3} 
\nn
\\
&& \times
\int \frac{d^3\bk^{(2)}}{(2\pi)^3}  \frac{d^3\bk^{(4)}}{ (2\pi)^3}
G[ \bar \psi_+ (\bl^{(3)};\bv_F^{(3)}) t^a \psi_+ (\bl^{(1)};\bv_F^{(1)}) ]
\nn
\\
&&\times
[ \bar \Psi_{+} (\bk^{(4)}) t^{a} \Psi_{+} (\bk^{(2)}) ]
.
\label{Effective_Int_QCDKondo}
\end{eqnarray}
Plugging the scaling dimensions of the fields and of the momenta discussed above, 
we find that the light-heavy four-Fermi operator has a {\it marginal} scaling dimension 
($ [dt]+2[d\ell_\para]+ 2 [\psi_+] = -1 + 2 - 1 = 0$) \cite{HIO}.

This result suggests that the four-Fermi interaction acquires logarithmic quantum corrections from 
the scattering of the light quark off the heavy-quark impurity with loop diagrams. 
We will see how the logarithmic enhancement arises 
from the second-order heavy-light scattering amplitude 
and determine the sign of the logarithmic correction. 
The following computation by the HD-EFT and HQ-EFT confirms the result in Ref.~\cite{Hattori:2015hka}.

Note again that the lower scaling dimension of $ \psi_+ $ 
due to the effective dimensional reduction is crucial for the Kondo effect. 
It is worth mentioning that the BCS instability, leading to superconductivity, can be understood as 
a consequence of the same dimensional reduction (see, e.g., Refs.~\cite{Polchinski:1992ed,Hong:1999ru, Hong:1998tn}). 
The RG analyses were performed for color superconductivity~\cite{Evans:1998nf, Evans:1998ek, 
Son:1998uk, Schafer:1998na, Hsu:1999mp}. 
Also, the scaling argument can be applied to the low-energy dynamics 
in a strong magnetic field where an analogous effective dimensional reduction occurs in the lowest Landau level \cite{HIO}. 
Consequences of this dimensional reduction are known as 
the magnetic catalysis of chiral symmetry breaking \cite{Gusynin:1994xp, Fukushima:2012xw, Hattori:2017qio} 
and the magnetically induced QCD Kondo effect \cite{Ozaki:2015sya}.

\subsection{Effective interaction and the leading-order scattering amplitude}

The gluon propagator at high density is split into two transverse structures 
\begin{eqnarray}
D^{\mu\nu}(k) = \frac{ i P_L^{\mu\nu}}{k^2 - \Pi_L} + \frac{ i P_T^{\mu\nu}}{k^2-\Pi_T}
- \xi \frac{k^\mu k^\nu}{k^4}
\, ,
\end{eqnarray}
where $ \xi $ is a gauge parameter 
and the diagonal color structure is suppressed for notational simplicity. 
The longitudinal and transverse projections are given by 
\begin{subequations}
\begin{eqnarray}
P_L^{\mu\nu} &=& - \left(g^{\mu\nu} - \frac{k^\mu k^\nu}{k ^2} \right) - P_T^{\mu\nu}
\, ,
\\
P_T^{\mu\nu} &=& \delta^{\mu i} \delta^{\nu j} 
\left(\delta^{ij} - \frac{k^i k^j}{\vert \bk \vert^2} \right)
\, .
\end{eqnarray}
\end{subequations}
In this section, we consider a normal phase (without Cooper pairing) and use the gluon self-energy 
from the hard dense loop approximation~\cite{Blaizot:1993bb, Manuel:1995td, Schafer:2003jn}. 
In the heavy-quark limit, we only need the electric component which is screened 
by the Debye screening mass $ \Pi_L \to m_D^2 = N_f (g\mu)^2/(4\pi^2) $ in the static limit.

We consider a light quark scattering off a static heavy quark, 
and identify the effective coupling $  G$ in Eq.~(\ref{Effective_Int_QCDKondo}) with 
the $S$-wave projection of the color electric interaction~\cite{Ozaki:2015sya,HIO}
\begin{eqnarray}
- i G 
&\equiv& (ig)^{2} \frac{1}{2}  \int^{1}_{-1} d \, {\rm{cos}} \theta \  D_{00} (k ) 
\nonumber \\
&\simeq& -i \frac{g^2}{2}  \int^{1}_{-1} 
 \frac{d \, {\rm{cos}} \theta }{  2 \mu^{2} ( 1 - {\rm{cos}} \theta ) + m_{D}^{2} } 
 \nonumber \\
&\simeq& -i \frac{ g^{2} }{ 4 \mu^{2} } {\rm{log}} \left( \frac{ 4 \mu^{2} }{ m_{\rm{D}}^{2} } \right)
\, .
\label{eq:G-Kondo}
\end{eqnarray}
We have put the initial and final momenta of the light quark on the Fermi surface 
since the Kondo effect occurs in such a low-energy regime. 
This leads to a spacelike momentum transfer $ k^2=(p^{(3)} - p^{(1)} )^2 = -2\mu^2 (1-\cos \theta) $, 
and we have integrated out the scattering angle $ \theta \equiv \arccos (\bm{v}_{F}^{(1)} \cdot \bm{v}_{F}^{(3)} /\mu^2)$. 
The gauge term proportional to $ \xi $ is suppressed in this kinematics. 
A similar $  S$-wave projection was performed for the Cooper pairing~\cite{Hsu:1999mp, Son:1998uk, Evans:1998ek}.

Using the effective coupling (\ref{eq:G-Kondo}), 
the leading-order scattering amplitude is given by 
\begin{eqnarray}
i \mathcal{M}^{(1)}
&=& - i G \sum_{r=1}^{N_c^2-1} (t^{r})_{i j} (t^{r})_{l m}
\label{eq:M0-Kondo}
\, ,
\end{eqnarray}
where $  N_c$ is the number of colors. 
For a notational simplicity, we have suppressed the spinor structure, $ [\bar u_+ \gam^0 u_+][ \bar U_+ U_+] $ 
with the projected spinors $ u_+ = \prj_+ u$ and $ U_+ = \Q_+ U $ for the light and heavy quarks, respectively.

\subsection{Kondo scale emerging from the RG evolution}

Since the four-Fermi operator, composed of light and heavy quarks, has a marginal scaling dimension, 
we anticipate emergence of a dynamical infrared scale. 
Below, we shall see how the logarithmic correction from the loop integral 
drives the RG evolution of the effective coupling to an infrared Landau pole. 

\begin{figure}[t]
     \begin{center}
              \includegraphics[width=\hsize]{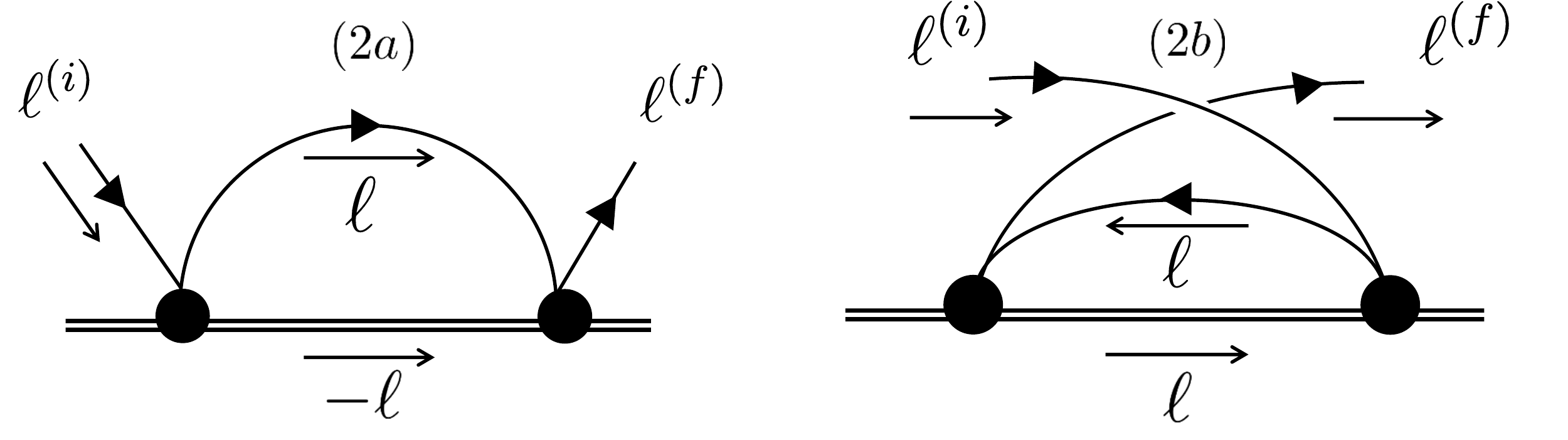}
     \end{center}
\vspace{-.5cm}
\caption{One-loop diagrams that give rise to logarithmic corrections. 
The single and double lines denote the light and heavy quarks. 
Both initial and final light-quark momenta are put on the Fermi surface, i.e., $ \ell_{(i)}^\mu = \ell_{(f)}^\mu =0 $. 
}
\label{fig:one-loop}
\end{figure}

At the one-loop level, there are two relevant diagrams for the Kondo effect (cf., Fig.~\ref{fig:one-loop}). 
In terms of the effective coupling $G$, the propagators (\ref{eq:S_HD}) and (\ref{eq:Qprop}), 
those amplitudes are written down as 
\begin{subequations}
\label{amplitude_Kondo}
\begin{eqnarray}
i \mathcal{M}^{(2a)}
&=& 
(-1) G^2 \T^{(a)} \sum_{\bv_F} \int \!\! \frac{d^4\ell}{(2\pi)^4}
\nn
\\
&&\times
\left[ \, \bar u_+  \gam^0 S (\ell) \gam^0  u_+ \, \right] 
\left[ \, \bar U_+   S_H(-\ell)   U_+ \, \right] 
\nn
\\
&=&
\sum_{\bm{v}_{F}}  \int \frac{ d^{4} \ell }{ (2\pi)^{4}} 
\frac{ 
G^{2} \mathcal{T}^{\rm{(a)}}  }
{( l^{0} - \ell_{\parallel}  + i \ell^0 \epsilon )  ( - \ell^{0} + i \epsilon )} 
\nn
\, ,
\\
\label{amplitude_a_Kondo}
\\
i \mathcal{M}^{(2b)}
&=& 
(-1) G^2 \T^{(b)} \sum_{\bv_F} \int \!\! \frac{d^4\ell}{(2\pi)^4}
\nn
\\
&&\times
\left[ \, \bar u_+  \gam^0 S (\ell) \gam^0  u_+ \, \right] 
\left[ \, \bar U_+   S_H(\ell)   U_+ \, \right] 
\nn
\\
&=& 
\sum_{\bm{v}_{F}}  \int \frac{ d^{4} \ell }{ (2\pi)^{4}} 
\frac{ 
G^{2} \mathcal{T}^{\rm{(b)}}  }
{( \ell^{0} - \ell_{\parallel}  + i  \ell^0 \epsilon )  (  \ell^{0} + i \epsilon ) } 
\nn
\, ,
\\
\label{amplitude_b_Kondo}
\end{eqnarray}
\end{subequations}
where the overall minus signs come from the Fermionic statistics 
and $ S_H $ is the heavy-quark propagator given in Eq.~(\ref{eq:Qprop}). 
Again, we put the initial and final momenta of the light quark on the Fermi surface, 
i.e., $\ell^{(i)}_\mu = \ell^{(f)}_\mu = 0  $, 
and suppressed the spinor structures which are found to be the same as that of the tree-level amplitude (\ref{eq:M0-Kondo}). 
We assume a static heavy quark with a vanishing velocity. 
As we will see shortly, it is important to have noncommutative color matrices 
\begin{subequations}
\begin{eqnarray}
\mathcal{T}^{({\rm{a}})}_{i j; lm} &=& 
\sum_{s,r} \sum_{k, k^{\prime} } (t^{r})_{i k } (t^{s})_{k j} (t^{r})_{l k^{\prime}} (t^{s})_{k^{\prime} m }
\, ,
\\
\mathcal{T}^{\rm{(b)}}_{ij ; lm} &=& 
\sum_{s,r}  \sum_{k k^{\prime}} (t^{r})_{i k } (t^{s})_{k j} (t^{s})_{ l k^{\prime}} (t^{r})_{k^{\prime} m }
\, .
\end{eqnarray}
\end{subequations}
Performing the integrals in Eq.~(\ref{amplitude_Kondo}), we have
\begin{subequations}
\label{amplitude_Kondo2}
\begin{eqnarray}
 \mathcal{M}^{(2a)}
&=& G^{2} \mathcal{T}^{\rm{(a)}} \, \rho_{\rm{F}}  
\int \frac{ d \ell_{\parallel} }{ \ell_{\parallel} } \theta(\ell_\para)
\label{amplitude_a_Kondo2}
\, ,
\\
 \mathcal{M}^{(2b)}
&=&  G^{2} \mathcal{T}^{\rm{(b)}}\, \rho_{\rm{F}}  
\int \frac{ d \ell_{\parallel} }{ \ell_{\parallel} } \theta(-\ell_\para)
\, .
\label{amplitude_b_Kondo2}
\end{eqnarray}
\end{subequations}
Note that, since the integral regions in Eq.~(\ref{amplitude_Kondo2}) are restricted 
to the above and below of the Fermi surface, 
Diagram (2a) and (2b) of Fig.~\ref{fig:one-loop} provide a particle and hole contribution, respectively, 
as specified by the pole positions in Eq.~(\ref{amplitude_Kondo}). 
The density of states on the Fermi surface $\rho_{\rm{F}}$ has been obtained from its area: 
\begin{eqnarray}
\rho_{\rm{F}}
= \sum_{\bm{v}_{F}} \int \frac{ d^{2} \ell_{\perp} }{ (2\pi)^3 } 
= \frac{ \mu^{2} }{ 2\pi^{2} }
\, .
\end{eqnarray}

We now examine an increment when the energy scale $ \Lam $ is reduced to $ \Lam - d\Lam $.  
The sum of the two one-loop amplitudes, integrated over a thin shell of a thickness $ d\Lam $, is obtained as
\begin{eqnarray}
\mathcal{M}^{(2)}
&=&  \mathcal{M}^{(2a)} + \mathcal{M}^{(2b)}
\nonumber 
\\
&=&  G^{2}  \rho_{\rm{F}} \, {\rm{log}} \left( \frac{ \Lambda }{ \Lambda - d \Lambda} \right) 
(\mathcal{T}^{\rm{(a)}}-\mathcal{T}^{\rm{(b)}}) 
\, .
\end{eqnarray}
The relative minus sign in the curly brackets originates from 
the fact that the particle and hole contributions in Eq.~(\ref{amplitude_Kondo2}) have opposite signs. 
The logarithms from the two distinct diagrams would cancel each other if the interaction were an Abelian type. 
Therefore, the non-Abelian nature of the interaction plays an essential role 
for the logarithmic correction to survive in the total amplitude. 
By the use of an identify $ \mathcal{T}^{({\rm{a}})}_{ij; lm} - \mathcal{T}^{({\rm{b}})}_{ij; lm} 
= - N_c/2 \sum_r (t^{r})_{ij} (t^{r})_{lm}  $, 
we find a logarithmic correction to the total one-loop amplitude 
\begin{eqnarray}
\mathcal{M}^{(2)}
&=&  G^{2} \frac{ N_{c} }{ 2 } \, \rho_{\rm{F}} \, {\rm{log}} \left( \frac{ \Lambda }{ \Lambda - d \Lambda} \right) 
\sum_r (t^{r})_{ij} (t^{r})_{lm} 
 .
\label{eq:M1_total}
\end{eqnarray}
It is this logarithm that renormalizes the effective coupling $  G$.

Now, combining the results in Eq.~(\ref{eq:M0-Kondo}) and (\ref{eq:M1_total}), 
we obtain the RG equation 
\begin{eqnarray}
\Lambda \frac{ d G }{ d \Lambda }
= - \frac{ N_{c} }{ 2 } \rho_{\rm{F}} G^{2}.
\end{eqnarray}
The solution to this RG equation is found to be 
\begin{eqnarray}
G(\Lambda)
= \frac{ G (\Lambda_0) }{ 1 + 2^{-1}  N_{c} \rho_{\rm{F}} G(\Lambda_0) 
{\rm{log}} (\Lambda /\Lambda_0) }
\, .
\label{Solution_QCDKondo}
\end{eqnarray}
Here, $\Lambda_0$ is the initial energy scale, 
and the initial condition of $G$ is given by the tree-level result in Eq.~(\ref{eq:G-Kondo}). 
Namely, $G(\Lambda_0)=( g^{2}/4\mu^{2} ){\rm{log}} ( 4\mu^{2} / m_{D}^{2} )$. 
It is clear that the effective coupling (\ref{Solution_QCDKondo}) is enhanced 
according to a negative beta function, when $ G (\Lambda_0) > 0 $. 
We can read off the location of the Landau pole that is called the Kondo scale \cite{Hattori:2015hka, HIO}: 
\begin{eqnarray}
\Lambda_{\rm{K}}
= \mu \, {\rm{exp}} \left( - \frac{ 2 }{ N_{c} \rho_{\rm{F}} G(\Lambda_0) } \right) 
,
\end{eqnarray}
where we took the initial energy scale at the hard scale $ \Lambda_0 = \mu $. 
When the temperature is reduced below the Kondo scale, the system becomes nonperturbative 
no matter how small the initial coupling $G(\Lambda_0)$ or $\alpha_{s}$ is.

Intuitively speaking, the light particles (or carriers of transport phenomena) are trapped around the impurity 
due to the strong-coupling nature of the low-energy dynamics. 
As a consequence, the (electrical) resistance is enhanced below the Kondo temperature $  T_K $ 
of which the scale is given by $ T_K \sim \Lambda_K $, and there emerges a minima at $  T_K$. 

%
%
%
\section{2SC Kondo effect}

\label{sec:2SC}

We here highlight four important ingredients for the Kondo effect discussed in the last section. 
First of all, the Kondo effect needs (i) {\it impurities}. 
Next, as implied by the essential degrees of freedom in HD-EFT and the scaling argument, 
the (1+1)-dimensional dispersion relation plays a crucial role. 
Therefore, the Kondo effect needs (ii) the existence of the Fermi surface 
so that the {\it effective dimensional reduction} occurs in the low-energy dynamics. 
For this low dimensionality, the four-Fermi operator, 
composed of the heavy and light particles, acquires a {\it marginal} scaling dimension. 
Consequently, the effective coupling strength is renormalized 
due to (iii) the {\it logarithmic quantum correction} from the loop integrals. 
Finally, the logarithms from the two distinct one-loop diagrams do not cancel out 
(iv) only when the interaction is a {\it non-Abelian} type.

Once these ingredients are identified, one may consider extensions of the Kondo effect. 
As already discussed, the QCD Kondo effect in dense quark matter is a straightforward extension 
since the fourth ingredient (iv) is provided by the color exchange interaction \cite{Hattori:2015hka}. 
One may also replace the second ingredient (ii), the Fermi surface, by a strong external magnetic field 
which also causes an effective dimensional reduction in the low-lying state, i.e., the lowest Landau level \cite{Ozaki:2015sya}. 

In this section, we propose a Kondo effect which occurs without impurities. 
Namely, we do not introduce the most essential ingredient (i) externally, 
but investigate 
a situation in which the gapped excitations (i.e., the ``impurities") emerge 
dynamically through a spontaneous symmetry breaking. 
The physical system we will consider is the 2SC phase of dense quark matter. 
The gapped quarks and the broken generators of the color symmetry 
will play a role of a heavy impurity and a non-Abelian interaction, respectively. 
We anticipate that a novel Kondo effect emerges 
with all the necessary ingredients inherent in the 2SC phase.

\begin{figure}
     \begin{center}
              \includegraphics[width=0.9\hsize]{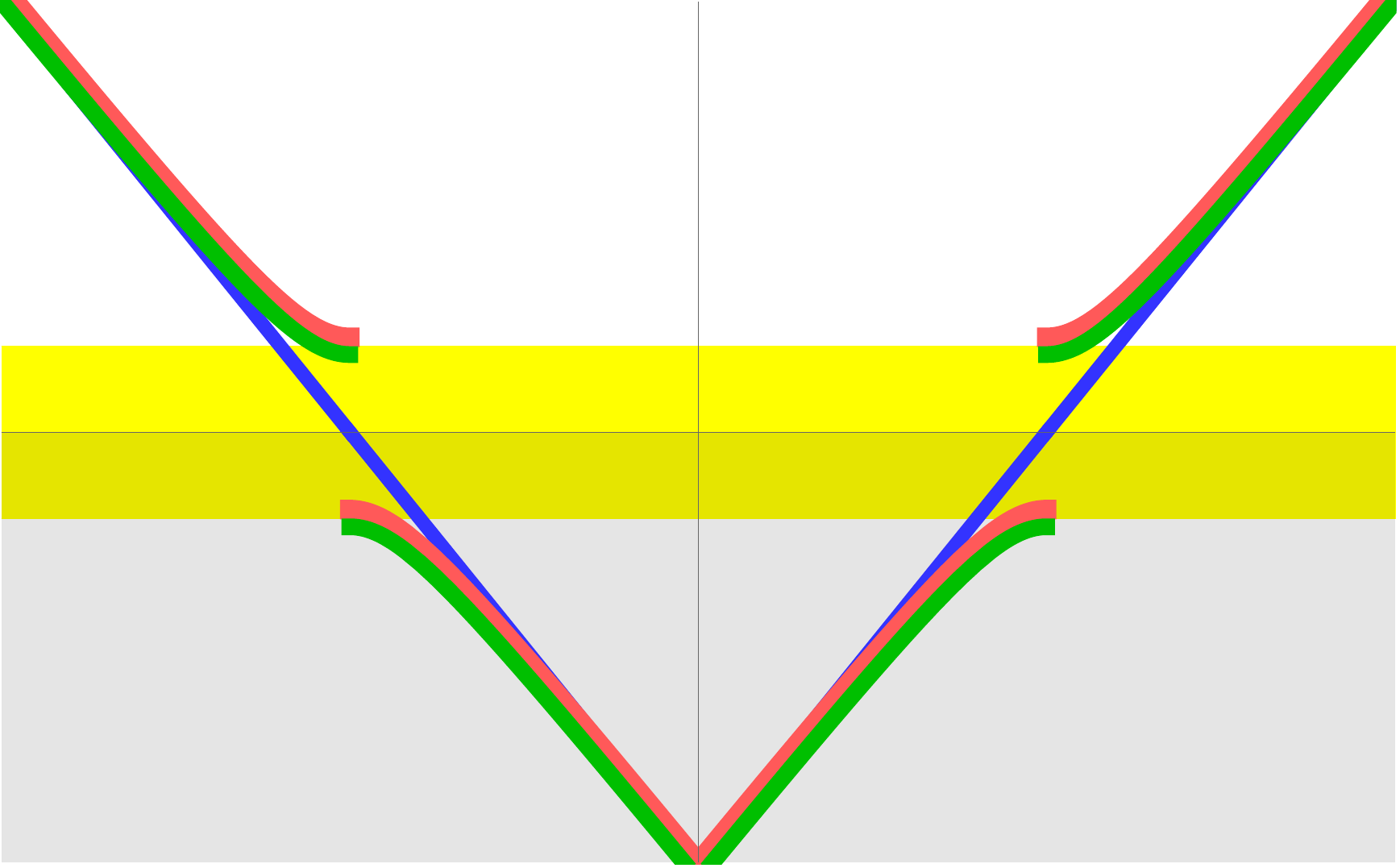}
     \end{center}
\caption{Dispersion relations of the gapped quarks (red and green curves) 
and ungapped quark (blue curve) near the Fermi surface. 
The ungapped quark has a finite density of states in the gapped region of the other quarks (filled yellow band).
}
\label{fig:dispersions}
\end{figure}

\subsection{Gapped quarks}

We have examined the quark propagator in normal phase in the previous section. 
Here, we prepare a quark propagator for a gapped quark. 
Without losing generality, we hereafter choose blue quarks ($ i=3 $) to be ungapped ones, 
so that red and green quarks are gapped above the Fermi surface (cf., Fig.~\ref{fig:dispersions}). 
Then, the color structure of the quark propagator reads 
\begin{eqnarray}
S_{fg}^{ij} 
&=&\delta_{fg} \left\{ \, (\delta^{ij}-\delta^{i3}\delta^{j3} ) S_\Delta + \delta^{i3}\delta^{j3}  S \, \right\}
\label{eq:quark_prop-2SC}
\, ,
\end{eqnarray}
where $ S $ and $ S_\Delta $ are the quark propagators with and without an energy gap, respectively. 
There would be off-diagonal components in the flavor space ($f,g $) 
if one considers interactions between the quasiexcitations and the $ \langle u d \rangle $ condensate in the 2SC phase. 
However, they are suppressed with a small value of the condensate. 
As shown in Ref.~\cite{Pisarski:1999av,Rischke:2000qz}, 
the propagator of the gapped quasiexcitations is given by 
\begin{eqnarray}
S_\Delta (p) &=& i \frac{p^0 - (\mu-\vert \bp \vert)}{ (p^0)^2-\ep_\bp^2 + i \epsilon} \prj_+ \gam^0
+ i \frac{p^0 - (\mu + \vert \bp \vert)}{ (p^0)^2- \bar \ep_\bp^2 + i \epsilon} \prj_- \gam^0
.
\nn
\\
\end{eqnarray}
The dispersion relations read 
\begin{eqnarray}
\ep_\bp = \sqrt{ ( \vert \bp\vert - \mu)^2 + \Delta^2 }
\, , \quad
\bar \ep_\bp = \sqrt{ ( \vert \bp\vert + \mu)^2 +  \bar \Delta^2 }
\, ,
\nn
\\
\end{eqnarray}
where the energy gaps $\Delta  $ and $ \bar  \Delta $ are generated as a consequence 
of the di-quark and di-aniquark condensate formation, respectively. 
As in Eq.~(\ref{eq:mom_decomp}) for the HD-EFT, we decompose 
the momentum into a large Fermi momentum and a small fluctuation near the Fermi surface. 
Then, we have 
$\vert \bp \vert ^2 \sim \mu^2 + 2\mu \ell_\para  $. 
Therefore, we find  
$ \vert \bp \vert \sim \mu + \ell_\para $ 
and $ \ep_\bp \sim \sqrt{ \ell_\para^2 + \Delta^2} =: \ep_\ell$. 
At the leading order in the $ 1/\mu$ expansion, the propagator reads 
\begin{eqnarray}
S_\Delta(p) \simeq 
\frac{i(\ell^0 + \ell_\para)}{ \ell_0^2 - \epsilon_\ell^2 +  i \vep}
\prj_+ \gam^0
\, .
\end{eqnarray}
The imaginary displacements are explicitly shown for 
the quasiparticle and quasihole excitations. 
This propagator has the same projection operator $\prj_+  $ as 
that of the ungapped quark (\ref{eq:S_HD}), 
meaning that the highly suppressed antiparticle excitations are neglected 
and that the coupling to a gluon field is again simplified 
as in the leading-order Lagrangian of the HD-EFT (\ref{eq:ph}). 
Namely, the gamma matrix is replaced by the Fermi velocity $v_F^\mu  $.

%
%

\subsection{Two effective coupling strengths from screened gluon exchanges}

The diquark condensation modifies not only the quark dispersion relation 
but also properties of the gluons. 
In the 2SC phase, the $ SU(3) $ color symmetry is broken down to the $ SU(2) $ subgroup, 
and the three gluons remain massless and do not interact with the ungapped blue quarks. 
They are also decoupled from the other two gapped quarks in the low-energy region 
below the size of the gap (cf., the yellow band in Fig.~\ref{fig:dispersions}). 
Therefore, a pure gluodynamics of $ SU(2) $ theory was investigated in such a low-energy region \cite{Rischke:2000cn}. 
Here, we consider the other five gluons which are screened by the Meissner mass $[ \Pi_T(0)=m_M^2 ]$ 
as well as the Debye mass $[ \Pi_L(0)=m_D^2] $ 
\cite{Rischke:2000qz, Casalbuoni:2001ha}.\footnote{
The gluon self-energy acquires off-diagonal elements. 
One can, however, diagonalize the self-energy by a unitary matrix, 
so that we assume that the color indices have been diagonalized \cite{Rischke:2000qz}.} 
Those gluons mediate the interactions between the gapped and ungapped quarks.

As we have seen in the last section, the energy gap of an impurity should be a dominant scale 
so that an impurity field is invariant under the scaling and plays the role of heavy scattering center. 
In the current case, the gap size $ \Delta $ should be the dominant scale 
for a gapped-quark field to be invariant under the scaling of an ungapped-quark field. 
This is realized in a scattering between a pair of gapped and ungapped quarks 
moving in almost the same direction within a relative angle $ \Delta/\mu $, 
where the momentum scale of ungapped quark is smaller than $ \Delta $ 
in the comoving frame of the gapped quark. 
Therefore, we focus on those pairs of which the incoming momenta 
are labelled by the same Fermi velocity $ \bv_F $. 
Then, dynamics is dominated by small fluctuations near the Fermi surface.

Since the gamma matrix on the interaction vertex is replaced by the four Fermi velocity $ v_F^\mu $, 
we identify the effective coupling strength from the $ S $-wave scattering [cf., Eq.~(\ref{eq:G-Kondo})]
\begin{eqnarray}
G
&=& i (ig)^{2} \frac{1}{2}  \int^{1}_{-1} d \cos\theta \ 
 v_F^{(1)\mu} v_F^{(2) \nu} D_{\mu\nu} (k ) 
\nonumber \\
&\simeq& \frac{g^2}{2}  \int^{1}_{-1} \!\! dz  
\left[  \frac{1}{  2 \mu^{2} ( 1 -z ) + m_{D}^{2} } -  \frac{(2-z)z }{  2 \mu^{2} ( 1 - z) + m_{M}^{2} }  \right]
\nn
\\&\simeq& \frac{ g^{2} }{ 2 \mu^{2} }\left( 1 +  \log  \frac{ m_M }{ m_D }  \right)
\, .
\label{eq:G-2SC}
\end{eqnarray}
As in the previous section, we have put the initial and final momenta of the ungapped quark on the Fermi surface, 
and integrated out the scattering angle $ \theta $. 


The magnitudes of $ m_D $ and $m_M  $ depend on the color index of gluons. 
Since the red and green quarks play the same role, 
we consider the scattering between the red (1) and blue (3) quarks. 
Then, there are only three relevant gluons $ A^{4,5,8} $ 
which mediate the interactions between the red and blue quarks.\footnote{ 
We employ the convention of the Gell-Mann matrices given in Sec.~15 of Ref.~\cite{PS}. 
Note also that the $ A^{4,5,8} $ do not cause mixing between the gapped quarks. 
Thus, the green quarks do not appear in the intermediate states 
of the one-loop scattering diagrams for the red and blue quarks, 
when interactions with the condensate can be neglected in Eq.~(\ref{eq:quark_prop-2SC}). 
} 
Quoting the results in Table~I of Ref.~\cite{Rischke:2000qz}, 
we have $m_M/m_D = 1/\sqrt{ 3}  $ for $ A^{4,5} $  and $ m_M/m_D = 1/3  $ for $ A^{8} $ in the 2SC phase at $  T=0$. 
Therefore, we define two effective coupling strengths 
\begin{subequations}
\label{eq:G-def}
\begin{eqnarray}
\label{eq:G1}
G_\SC  &=&  \frac{g^2}{2\mu^2} (1- \frac{1}{2} \log 3) > 0
\, ,
\\
\label{eq:G2}
\bar G_\SC &=& \frac{g^2}{2\mu^2}  (1-\log3)  < 0
\, .
\end{eqnarray}
\end{subequations}
It is useful to compare the setup in the 2SC phase to the anisotropic Kondo effect discussed in Ref.~\cite{anderson1970poor, Anderson:1970}. Our coupling strengths $G_\SC$ and $\bar G_\SC$ play 
similar roles of $J_\pm$ and $J_z$ there. 
A negative $J_z$ corresponds to the ferromagnetic Kondo problem 
of which the fate at low-energy depends on the relative strength between $J_\pm$ and $|J_z|$. 
In order to determine the low-energy physics, 
we will investigate how those effective couplings $G_\SC$ and $\bar G_\SC$ are renormalized
with the next-to-leading order scattering processes between gapped and ungapped excitations.


%
%
%
%

\subsection{Color flows in the next-to-leading order scattering diagrams}

\begin{figure*}[t]
     \begin{center}
              \includegraphics[width=0.75\hsize]{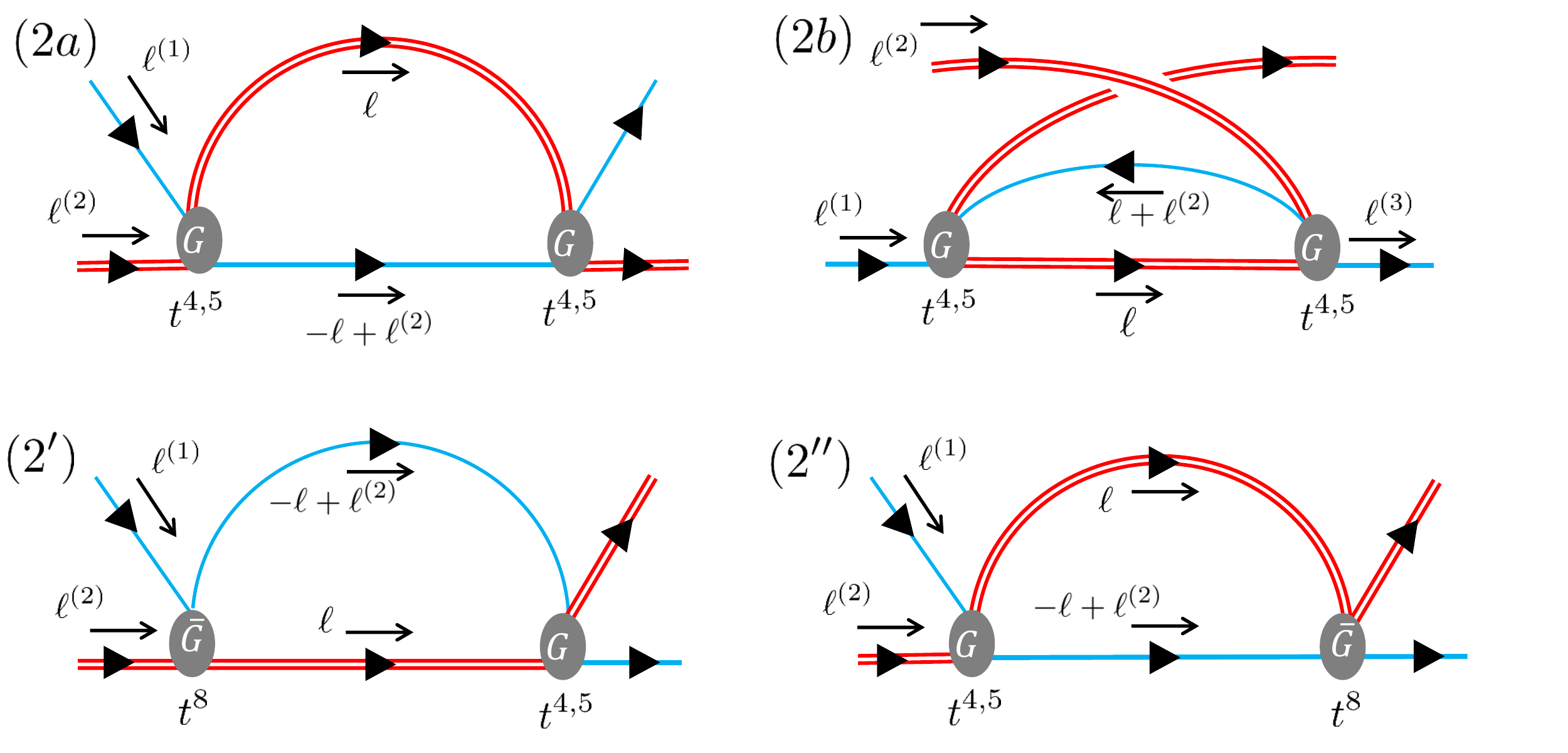}
     \end{center}
\vspace{-0.5cm}
\caption{Next-to-leading order scattering diagrams which renormalize the effective coupling constants. 
Single and double lines show ungapped and gapped quarks, respectively.}
\label{fig:diagrams_2SC}
\end{figure*}

We write the scattering diagrams in terms of the effective coupling constants defined above. 
At the leading order, we simply have 
\begin{eqnarray}
i \M^{(1)}_\SC &=& -i G_\SC \sum_{a=4,5}
\left[ \, \bar u_+^3  \gam^0 (t^a)^{31}  u_+^1 \, \right] \left[ \, \bar u_+^1  \gam^0 (t^a)^{13}   u_+^3 \, \right] 
\nn
\\
&& -i
\bar G_\SC \left[ \, \bar u_+^1  \gam^0 (t^8)^{11}  u_+^1 \, \right]  \left[ \, \bar u_+^3  \gam^0 (t^8)^{33}   u_+^3 \, \right] 
\nn
\\
&=&  -i \left(\,
\frac{  G_\SC}{2} \U^{31} \U^{13}
- \frac{ \bar G_\SC}{6}  \U^{11} \U^{33}
\, \right)
\label{eq:M0}
,
\end{eqnarray}
where we defined a shorthand spinor notation $ \U^{ab} \equiv \bar u_+^a  \gam^0  u_+^b $.

Next, we shall identify next-to-leading order diagrams which renormalize the effective coupling strengths. 
In Fig.~\ref{fig:diagrams_2SC}, we show flows of color charges 
where the color matrices $ t^{4,5} $ mix the red and blue quarks, 
while the diagonal color matrix $ t^8 $ does not. 
Similar to the diagrams in Fig.~\ref{fig:one-loop}, 
we should consider a pair of diagrams $ (2a) $ and $(2b)  $ of Fig.~\ref{fig:diagrams_2SC}. 
However, we do not need to consider the diagrams 
with the diagonal matrix $t^8  $ on all the vertices, 
since possible logarithmic corrections will cancel out. 
On the other hand, we should include Diagrams $ (2^\prime) $ and $ (2^{\prime\prime})  $ of Fig.~\ref{fig:diagrams_2SC} 
which have each of $ t^{4,5} $ and $t^8  $. 
Neither of these two diagrams has a relevant cross channel, 
since its cross channel is a disconnected diagram, indicating an annihilation 
between an ungapped particle and gapped hole, or an ungapped hole and gapped particle. 
Diagrams $ (2^\prime) $ and $ (2^{\prime\prime})  $ of Fig.~\ref{fig:diagrams_2SC} contain a factor of $ G_\SC \bar G_\SC $, 
and give rise to a mixing between the two coupling strengths. 

Similar to the previous section, the scattering amplitudes of 
Diagram $ (2a) $ and $(2b)  $ of Fig.~\ref{fig:diagrams_2SC} are written down as 
%
%
%
%
\begin{subequations}
\begin{eqnarray}
i \M^{(2a)} _\SC&=& - ( G_\SC)^2 (\T^{(a)}_\SC)_{33,11} \sum_{\bv_F} \int \!\! \frac{d^4\ell}{(2\pi)^4}
\nn
\\
&&\times
\left[ \, \bar u_+^3  \gam^0 S_\Delta (\ell) \gam^0  u_+^3 \, \right] 
\left[ \, \bar u_+^1  \gam^0 S(\ell^{(2)}-\ell) \gam^0  u_+^1 \, \right] 
\nn
\\
&=& - (G_\SC)^2 (\T^{(a)} _\SC )_{33,11} \U^{33} \U^{11} \rho_F I_+
\, ,
\\
i \M^{(2b)} _\SC
&=& - (  G_\SC)^2 (\T^{(b)}_\SC)_{33,11}   \sum_{\bv_F} \int \!\! \frac{d^4\ell}{(2\pi)^4}
\nn
\\
&&\times
\left[ \, \bar u_+^3  \gam^0 S_\Delta (\ell) \gam^0  u_+^3 \, \right] 
\left[ \, \bar u_+^1  \gam^0 S(\ell^{(2)}+\ell) \gam^0  u_+^1 \, \right] 
\nn
\\
&=& - ( G_\SC)^2 (\T^{(b)}_\SC)_{33,11}  \U^{33} \U^{11}  \rho_F I_-
\, .
\end{eqnarray}
\end{subequations}
%
%
%
%
On the external lines, we put all the spatial momenta and energies of 
ungapped quarks on the Fermi surface. 
Only one finite component of the external momentum is 
the energy of a gapped quark, i.e., $\ell^{(2)}_{\mu=0} = \ell^{(4)}_{\mu=0} = \Delta  $. 
The density of states at the Fermi surface $ \rho_F $ is obtained as in the previous section. 
The longitudinal components of the integrals are given by 
\begin{eqnarray}
I_\pm =  \int \!\! \frac{d^2\ell_\para}{2\pi}
\frac{i(\ell^0 + \ell_\para)}{ \ell_0^2 - \epsilon_\ell^2 +  i \vep}
\cdot
\frac{i}{( \Delta \mp \ell^0) \pm \ell_\para + i (\Delta \mp \ell^0 ) \vep}
\, .
\nn
\\
\end{eqnarray}
Here, we clearly see that $ I_+ = - I_- =: I_\para$. 
The structures of the color matrices read  
\begin{subequations}
\begin{eqnarray}
(\T^{(a)}_\SC)_{ij,mn}&=& (t_5 t_4)_{ij} (t_5 t_4)_{mn} +  (t_4 t_5)_{ij} (t_4 t_5)_{mn}
\, ,
\nonumber
\\
\\
(\T^{(b)}_\SC)_{ij, mn} &=&  (t_5 t_4)_{ij} (t_4 t_5)_{mn} +  (t_4 t_5)_{ij} (t_5 t_4)_{mn} 
\, .
\nn
\\
\end{eqnarray}
\end{subequations}
We find that $ ( \T^{(a)}_\SC)_{33,11} = - (\T^{(b)}_\SC)_{33,11} = 1/8 $. 
Summarizing those observations, one can write the sum of the two amplitudes as 
\begin{eqnarray}
i\M^{(2)} _\SC&\equiv& i \M^{(2a)} _\SC+ i \M^{(2a)}_\SC
\nn
\\
&=& -  \frac{1}{4}  (G_\SC)^2  \rho_F  I_\para  \, \U^{33} \U^{11}
\, .
\end{eqnarray}
Performing the contour integral with respect to $ \ell^0 $ 
and keeping only the singular terms when $ \ell_\para \to 0 $,\footnote{
One can enclose the contour either in the upper or lower half planes. 
The results are the same in both cases.} 
we have  
\begin{eqnarray}
I_\para \simeq - \frac{i}{2} \int   d \ell_\para  
\frac{ \theta(\ell_\para) }{\ell_\para + \frac{\ell_\para^2}{2\Delta} }
\, .
\end{eqnarray}
Integrating over a shin shell $ \Lam - d\Lam \leq \ell_\para \leq \Lam $, 
we find a logarithmic contribution 
\begin{eqnarray}
I_\para =  - \frac{i}{2}  \ln \frac{\Lam}{\Lam - d\Lam} +   \order (  \Lam) 
\, .
\end{eqnarray}
The subsequent terms are a polynomial of $ \Lam $ which provides only irrelevant corrections. 
Plugging this result into $ \M^{(2)} _\SC$, we obtain 
\begin{eqnarray}
\label{eq:M12}
\M^{(2)}_\SC = 
 \frac{1}{8}   ( G_\SC)^2  \rho_F   \ln \frac{\Lam}{\Lam - d\Lam}   \, \U^{33} \U^{11}
\, .
\end{eqnarray}

The remaining two diagrams can be computed in the same manner. 
Actually, those diagrams have the same kinematics as Diagram $ (2a) $ of Fig.~\ref{fig:diagrams_2SC}, 
so that we need only to take care of the color structures. 
Writing down those two contributions, we have 
\begin{subequations}
\begin{eqnarray}
i\M^{(2^\prime)} _\SC&=& 
 -  G_\SC \bar G_\SC (\T^{(c)}_\SC)_{13,31} \sum_{\bv_F} \int \!\! \frac{d^4\ell}{(2\pi)^4}
\nn
\\
&&\times
\left[ \, \bar u_+^3  \gam^0 S_\Delta (\ell) \gam^0  u_+^1 \, \right] 
\left[ \, \bar u_+^1  \gam^0 S(\ell^{(2)}-\ell) \gam^0  u_+^3 \, \right] 
\nn
\\
&=& -  G_\SC \bar G_\SC (\T^{(c)}_\SC)_{31,13}  \U^{31} \U^{13} \rho_F I_\para
\, ,
\\
i\M^{(2^{\prime\prime})} _\SC&=& 
 -  G_\SC \bar G_\SC (\T^{(d)}_\SC)_{13,31} \sum_{\bv_F} \int \!\! \frac{d^4\ell}{(2\pi)^4}
\nn
\\
&&\times
\left[ \, \bar u_+^1  \gam^0 S_\Delta (\ell) \gam^0  u_+^3 \, \right] 
\left[ \, \bar u_+^3  \gam^0 S(\ell^{(2)}-\ell) \gam^0  u_+^1 \, \right] 
\nn
\\
&=&-  G_\SC \bar G_\SC (\T^{(d)}_\SC)_{13,31}  \U^{13} \U^{31} \rho_F I_\para
\, ,
\end{eqnarray}
\end{subequations}
where the products of the color matrices read 
\begin{subequations}
\begin{eqnarray}
(\T^{(c)}_\SC)_{ij,mn}&=& (t_4 t_8)_{ij} (t_4 t_8)_{mn} +  (t_5 t_8)_{ij} (t_5 t_8)_{mn}
\, ,
\nn
\\
\\
(\T^{(d)}_\SC)_{ij,mn}&=& (t_8 t_4)_{ij} (t_8 t_4)_{mn} +  (t_8 t_5)_{ij} (t_8 t_5)_{mn}
\, .
\nn
\\
\end{eqnarray}
\end{subequations}
We find that $ (\T^{(c)}_\SC)_{31,13} =  (\T^{(d)}_\SC)_{13,31} = - 1/12 $. 
Therefore, Diagram $ (2^\prime) $ and $ (2^{\prime\prime}) $ of Fig.~\ref{fig:diagrams_2SC} provide the same contributions 
\begin{eqnarray}
\label{eq:M34}
\M^{(2^\prime)} _\SC= \M^{(2^{\prime\prime})}_\SC =
-  \frac{1}{24}  G_\SC \bar G_\SC \, \rho_F   \ln \frac{\Lam}{\Lam - d\Lam}   \, \U^{13} \U^{31}
\, .
\nn
\\
\end{eqnarray}

\subsection{Coupled RG equations and RG-flow diagram}

\begin{figure}
     \begin{center}
              \includegraphics[width=\hsize]{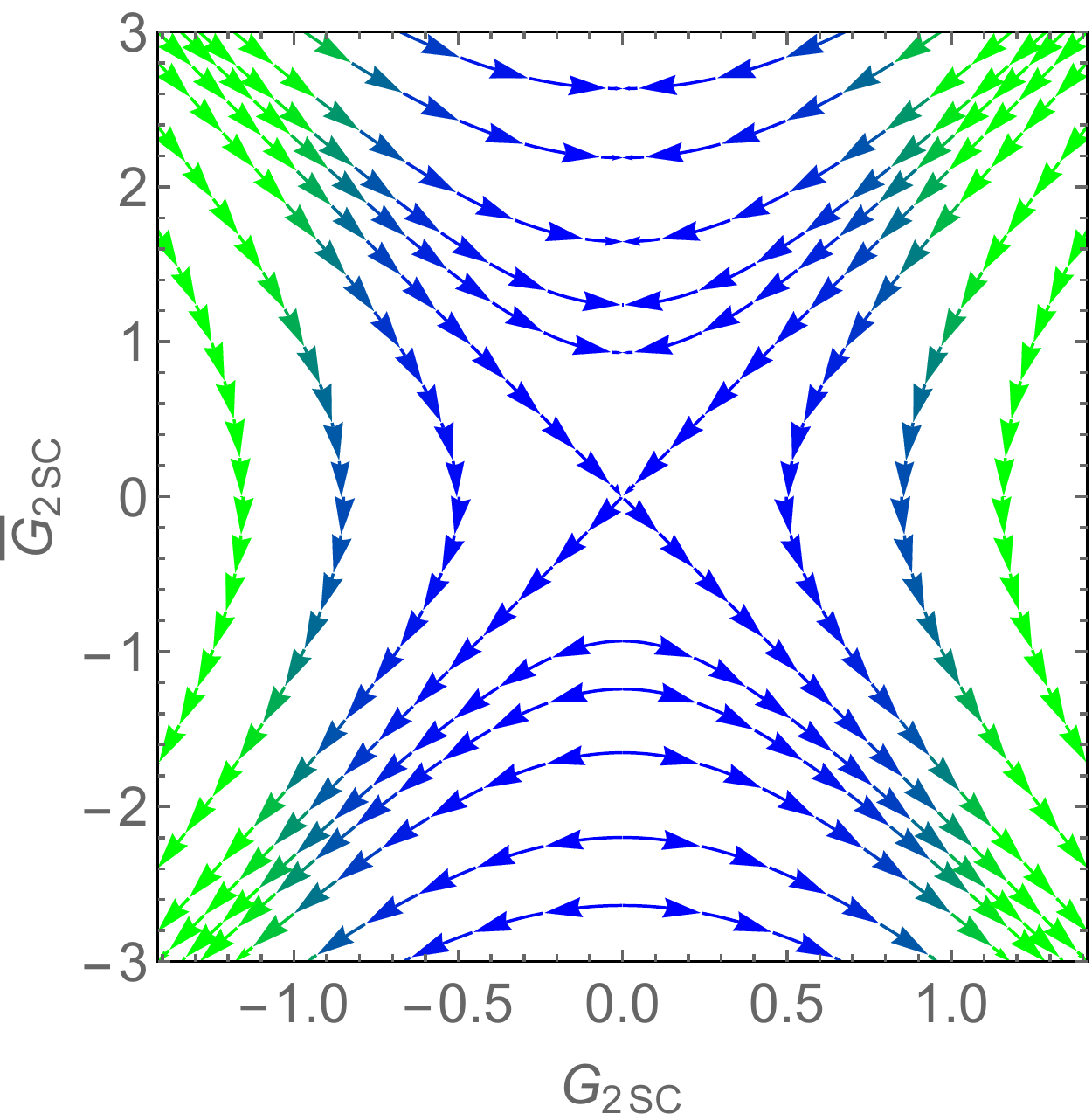}
     \end{center}
\vspace{-0.5cm}
\caption{RG flow in the direction of arrows along the parabolic curves.
The color scales from green to blue as the norm of the arrow increases. 
}
\label{fig:RG_flow}
\end{figure}

We are now in position to derive RG equations for $ G_\SC $ and $\bar G_\SC  $. 
Plugging the leading-order amplitude (\ref{eq:M0}) and 
the next-to-leading order amplitudes (\ref{eq:M12}) and (\ref{eq:M34}), 
we immediately obtain the coupled RG equations 
\begin{subequations}
\label{eq:RGs}
\begin{eqnarray}
\Lam \frac{d G_\SC}{d\Lam} &=& - \frac{1}{6}  \rho_F\, G_\SC \bar G_\SC
\, ,
\\
\Lam \frac{d  \bar G_\SC}{d\Lam} &=& -  \frac{3}{4}   \rho_F\, G_\SC ^2
\, .
\end{eqnarray}
\end{subequations}
Correspondingly, the right-hand sides of the RG equations provide two distinct beta functions. 
In Fig.~\ref{fig:RG_flow}, we draw the RG flow driven by the ``velocity field'' identified with those beta functions. 

To understand the RG-flow profile, we write the RG equations (\ref{eq:RGs}) as  
\begin{eqnarray}
\frac{d G_\SC}{d  \bar G_\SC} &=&  \frac{ 2}{9}  \cdot \frac{ \bar G_\SC }{ G_\SC }
\, .
\end{eqnarray}
This means that the RG flow evolves along parabolic curves 
\begin{eqnarray}
\label{eq:GGbar}
 (3G_\SC)^2 -  2  ( \bar G_\SC)^2 =  C
\, ,
\end{eqnarray}
where the constant $ C $ is determined by the initial conditions at $ \Lam = \Lam_0 $. 
We take the tree-level coupling strengths (\ref{eq:G-def}) as the initial conditions, 
and have a positive constant $ C>0 $. 
In Fig.~\ref{fig:RG_flow}, we start out at a point $ G_\SC(\Lam_0)>0 $ and $ \bar  G_\SC(\Lam_0) < 0 $, 
and find that the RG flow goes into the lower right corner. 
This means that the $ G_\SC(\Lam) $ and $ \bar G_\SC(\Lam) $ evolve 
toward positive and negative infinity, respectively, away from the weak-coupling regime near the origin. 
Thus, the RG evolution driven by the interaction between the gapped and ungapped quarks 
gives rise to a strongly coupled system in the low-energy regime. 
This is a characteristic behavior of the Kondo effect.

From the RG equations (\ref{eq:RGs}), we get 
\begin{eqnarray}
\Lam \frac{d  \bar G_\SC}{d\Lam} &=& -  \frac{1}{12}   \rho_F\, ( 2 \bar G_\SC ^2 + C)
\, ,
\end{eqnarray}
and its solution 
\begin{eqnarray}
\label{eq:sol}
\bar G_\SC(\Lam) = c \tan\left[ \,
 \frac{ c }{6} \rho_F \log \frac{\Lam_0}{\Lam} + \arctan\left( \frac{ \bar G_\SC(\Lam_0)}{ c} \right)
\,\right]
\, ,
\nn
\\
\end{eqnarray}
where $c = \sqrt{ C/2}  $. 
This solution hits a Landau pole when the argument of the tangent approaches $ \pi/2 $, 
giving rise to an emergent scale (see Ref.~\cite{Hattori:2017qio} for the same form of solution). 
According to the relation (\ref{eq:GGbar}), 
the other coupling strength $ G_\SC(\Lam) $ also hits the Landau pole at the same energy scale (cf., Fig.~\ref{fig:RG_flow}).

Noting that $ c \rho_F \sim {\mathcal O}(g^2) $ and $ \bar G_\SC(\Lam_0)/c \sim {\mathcal O}(1) $ in Eq.~(\ref{eq:sol}), 
the parametric dependence of the Kondo temperature is found to be 
\begin{eqnarray}
T_K^\SC = \Lam_0 \exp \left( \, - \frac{c_1}{g^2} \,\right)
\, .
\end{eqnarray}
The explicit form of the order-one constant $ c_1 $ can be obtained easily, 
but is suppressed for a simple parametric estimate. 
It is natural to take the initial scale at the gap size, $ \Lam_0 \sim \Delta \sim \mu \exp(-  c_2/g) $, 
where $ c_2 \sim  {\mathcal O}(1) $. 
Importantly, the gap size has a weaker exponential suppression because of the enhancement 
arising from the unscreened color-magnetic interaction \cite{Son:1998uk, Hsu:1999mp, Pisarski:1999bf, Pisarski:1999tv}. 
However, the gluons involved in the 2SC Kondo effect are screened by both the Debye and Meissner masses. 
Therefore, in the weak-coupling limit $ g \ll 1 $, we get a parametric estimate 
\begin{eqnarray}
\label{eq:scales}
T_K^\SC \sim \mu \exp \left( \, - \frac{c_1}{g^2} \,\right) \ll \Delta \ll \mu
\, .
\end{eqnarray}
This hierarchy confirms our basic picture of the 2SC Kondo effect (cf., Fig.\ref{fig:dispersions}). 
Note that the two dynamical scales $ T_K^\SC $ and $ \Delta $ emerge in the different color sectors.

The emergence of the strong-coupling regime implies formation of a bound state or condensation 
between the gapped and ungapped quarks. 
In condensed matter systems, such a bound state between the conduction electron and impurity has been known as the Kondo singlet  \cite{PhysRev.138.A1112, PhysRev.147.223, yamada2010electron}. 
While the impurity magnetic moment is localized in such systems, 
the gapped quarks are thermally excited in bulk. 
Therefore, in the present system, we may think of it as a condensate rather than a localized bound state. 
Then, the formation of condensation breaks the residual SU(2) color symmetry, 
and the associated gluons become massive via the Higgs mechanism.\footnote{
The formation of a bound state/condensate may give rise to a pole 
in the Bethe-Salpeter amplitude (cf., an analogous structure in the Cooper pairing \cite{abrikosov2012methods}), 
suggesting that the scattering ungapped quarks have off-shell momenta on the external lines. 
The natural scales of the off-shellness and binding energy are given by the Kondo scale (cf., Ref.~\cite{yamada2010electron}) 
which is the emergent and smallest scale in our analysis as shown in Eq.~(\ref{eq:scales}). 
The off-shellness or deviation from the Fermi surface below the emergent scale do not change 
the form of the logarithm that drives the renormalization flow from the ultraviolet to the Landau pole. 
This observation, {\it a postriori}, justifies our setup of the kinematics with the ungapped-quark momenta on the Fermi surface. 
}
Nevertheless, whether such a ``Higgs phase'' emerges depends on the gapped-quark distribution 
which reduces as we decrease temperature. 
At this moment, we conclude that the residual color symmetry is broken 
as long as the 2SC Kondo phase manifests itself in the QCD phase diagram. 
In more general, one may ask how a possible phase structure depends on the impurity distribution 
as an axis of an extended phase diagram (See Refs.~\cite{Yasui:2016svc, Yasui:2017izi} for the chiral symmetry breaking 
in the presence of a homogeneous distribution of the heavy-quark impurities in quark matter). 
We leave those issues as future works.

\section{Conclusions and discussions}
\label{sec:summary}

In this paper, we investigated the RG evolutions of the coupling strengths between 
gapped and ungapped quarks in the two-flavor superconducting (2SC) phase. 
The next-to-leading order diagrams generate logarithmic quantum corrections 
and have the effective coupling strengths renormalized. 
We obtained coupled RG equations for the two coupling strengths associated with distinct color channels. 
The RG-flow diagram indicates that both of the coupling strengths evolve into a strong-coupling regime 
as the energy scale is reduced toward the Fermi energy. 
This is a characteristic behavior of the Kondo effect, so that we call it the {\it 2SC Kondo effect}.

Once the system approaches the strong-coupling regime, 
the fate of the RG evolution needs to be investigated with nonperturbative methods. 
For example, a mapping from the Kondo problem in the vicinity of an infrared fixed point to conformal field theory 
has been known as a useful method (see recent works \cite{Kimura:2016zyv, Kimura:2018vxj} and references therein).
In the present system, one could ask if there is an infrared fixed point, 
and what the ground state of the system is.

The magnitudes of the Kondo effect on bulk quantities, such as transport coefficients, 
depend on the concentration of impurities, i.e., the Fermi-Dirac distribution of the gapped quarks in the present case. 
At strict zero temperature, there is no gapped quark excitations. 
Therefore, effects of the 2SC Kondo effect will be most prominent 
in between the transition temperature of the 2SC phase and zero temperature. 
This contrasts with the conventional Kondo effect which remains important at zero temperature.

If the 2SC phase exists in the neutron-star cores (see Ref.~\cite{Alford:2007xm}), 
we expect that the occurrence of the 2SC Kondo effect may have implications for the neutron-star physics. 
In particular, the transport properties of the neutron-star matter could be altered in a way similar to 
how the conventional Kondo effect affects the low-temperature resistivity of magnetic alloys. 
One typical example is the neutrino emissivity which is important as a cooling mechanism of neutron stars. 
It is conventionally known that, in the 2SC phase, the existence of 
ungapped quarks opens a large phase space for neutrino emission, 
leading to a too-fast cooling as compared to the astrophysical data~\cite{Jaikumar:2005hy}. 
The 2SC Kondo effect may serve as a mechanism to suppress the emissivity 
and give a cooling rate closer to the data. This calls for more detailed investigation in future.

Besides, we would like to mention a new possibility that ultracold atoms could serve 
as a designed platform for studying quantum many-body physics. 
Realization of ``color superconductivity'' has been discussed with fermionic atoms 
carrying color-like internal degrees of freedom \cite{PhysRevLett.98.160405, PhysRevLett.99.130406, 
Maeda:2009ev, PhysRevA.82.063615}. 
Those models, however, do not have ``gluons.'' 
Although the 2SC Kondo effect does not necessarily need dynamical gluon exchanges, 
which can be substituted by contact interactions, 
a non-Abelian matrix on each interaction vertex is an indispensable ingredient. 
While Cooper pairing with a ``color-flipping'' effect was recently discussed \cite{PhysRevA.97.023632}, 
a noncommutative property is yet more demanding. 
Nevertheless, the direction for ultracold atoms deserves further study. 
It may also be worth mentioning that realization of the Kondo effect \cite{PhysRevLett.111.215304, PhysRevLett.111.135301, PhysRevA.93.011606, PhysRevLett.121.203001} 
and Shiba state \cite{PhysRevA.83.033619,PhysRevA.83.061604,PhysRevA.83.063611}, 
a mixture of impurity and superconducting states, was discussed in terms of ultracold atoms.

%
%
\section*{acknowledgements}
The authors thank Sho Ozaki and Dirk Rischke for useful discussions. 
This work is partly supported by China Postdoctoral Science Foundation 
under Grants Nos.~2016M590312 and 2017T100266 (KH), 
the Young 1000 Talents Program of China and NSFC under Grants Nos.~11535012 and~11675041 (XGH). 
R.D.P. is funded by the U.S. Department of Energy for support under contract DE-SC0012704.

\appendix

\section{Effective field theories (EFTs)}

\label{sec:EFT}

In this appendix, we briefly summarize the basic properties of the high-density and heavy-quark effective field theories 
at the leading order of the expansions with respect to a large chemical potential $ 1/\mu $ and 
a heavy-quark mass $1/m_H  $, respectively.

\subsection{High-density EFT}

\label{sec:HDEFT}

For a given Fermi velocity, the corresponding plane wave is factorized as 
\begin{eqnarray}
\psi (x) = \sum_{\bv_F} e^{i \mu \bv_F \cdot \bx} 
\int_{\ell \ll \mu} \!\! \frac{d^4\ell}{(2\pi)^4} e^{-i\ell^\mu x_\mu} \psi_{\bv_F} (\ell)
\, ,
\end{eqnarray}
where $ \psi_{\bv_F} (\ell) $ is the field for low-energy excitations (in the momentum space). 
Plugging the above decomposition into the Lagrangian, one gets
\begin{eqnarray}
\Lag &=& \bar \psi(x) (i\sla D + \mu \gam^0) \psi(x)
\nonumber
\\
&=& \sum_{\bv_F} \bar \psi_{\bv_F} (x) ( i\sla D + \mu \sla v_{F+} ) \psi_{\bv_F} (x) 
\label{eq:HDEFT-kinetic}
\, ,
\end{eqnarray}
where $ v_{F\pm}^\mu \equiv (1, \pm \bv_F)$. 
The kinetic term in Eq~(\ref{eq:HDEFT-kinetic}) yields not only the one in Eq.~(\ref{eq:ph}) 
but also the other three terms 
\begin{subequations}
\begin{eqnarray}
&&
\bar \psi_- ( i\sla D + \sla v_{F+}^\mu ) \psi_- 
= \bar \psi_-  ( i v_{F-}^\mu D_\mu  + 2\mu )\gam^0  \psi_- 
\label{eq:anti}
\, ,
\\
&&
\bar \psi_\pm ( i\sla D + \sla v_{F+}^\mu ) \psi_\mp 
= \bar \psi_\pm ( i \gam_\perp^\mu D_\mu )  \psi_\mp 
\label{eq:mix}
\, ,
\end{eqnarray}
\end{subequations}
where $  \psi_\pm (x;\bv_F) \equiv \prj_\pm \psi(x;\bv_F) $, 
$\gam_\para^\mu \equiv (\gam^0, (\bv_F \cdot \bgam) \bv_F)$, 
and $\gam_\perp^\mu \equiv \gam^\mu - \gam_\para^\mu$. 
From Eq.~(\ref{eq:anti}), the antiparticle states are gapped by $ 2 \mu$, 
so that those excitations are highly suppressed in the dense system. 
When $ \ell_\perp $ is smaller than the gap $ 2\mu$ in Eq.~(\ref{eq:mix}), 
the mixing between the particle and antiparticle states is also suppressed.

Here are some basic properties of the projection operators: 
\begin{eqnarray}
&&
\prj_\pm \prj_\pm = \prj_\pm \, , \ \ \ 
\prj_\pm \prj_\mp = 0 \, , \ \ \ 
\prj_\pm^\dagger = \prj_\pm \, ,
\\
&&
\sla v_{F \pm} = 2 \gam^0 \prj_\mp \, , \ \ \ 
\prj_\pm \gam_\para^\mu = v_{F\mp}^\mu \gam^0 \prj_\mp \, , \ \ \ 
\prj_\pm \gam_\perp^\mu =  \gam_\perp^\mu \prj_\pm \, , 
\nn
\\
&&
\gam_{\perp \mu} v_{F\pm}^\mu = (\gam-\gam_\para)_\mu  v_{F\pm}^\mu
= \mp (\bgam \cdot \bv_F) (1-\bv_F^2) = 0
\nn
\, .
\end{eqnarray}
By using the identities, we find
\begin{subequations}
\begin{eqnarray}
\label{eq:prjs}
&&
\prj_\pm \gam^\mu \prj_\mp 
= v_{F \mp}^\mu  \prj_\pm \gam^0 \prj_\mp
\, ,
\\
&&
\prj_\pm \gam^\mu \prj_\pm 
= \prj_\pm \gam_\perp^\mu \prj_\pm
\, .
\end{eqnarray}
\end{subequations}

\subsection{Heavy-Quark EFT}

\label{sec:HQEFT}

We also briefly summarize the heavy quark effective field theory at the leading order \cite{Manohar:2000dt}. 
We shall decompose the heavy-quark momentum as 
\begin{eqnarray}
p^\mu = m_H v_H^\mu + k^\mu
\label{eq:p_Q}
\, ,
\end{eqnarray}
where the velocity is normalized as $ v_H^2 = 1$. 
Since excitations of the antiparticle states are highly suppressed by $ 1/m_H$, 
it is natural to introduce operators 
\begin{eqnarray}
\Q_\pm = (1 \pm \sla v_H )/2
\, ,
\end{eqnarray}
which, in the rest frame ($ \bv_H=0 $), project out the particle and antiparticle states. 
Here are some basic properties of the projection operators 
\begin{eqnarray}
\label{eq:Qid}
&&\hspace{-0.5cm}
\Q_\pm \Q_\pm 
= \Q_\pm
\, , \quad 
\Q_\pm \Q_\mp 
= 0
\, ,
\\
&&\hspace{-0.5cm}
\Q_\pm \gam^\mu \Q_\pm 
= \pm v_Q^\mu \Q_\pm
\, , \quad
\Q_\pm \gam^\mu \Q_\mp 
=  (\gam^\mu \pm v_Q^\mu) \Q _\mp
\nn
 .
\end{eqnarray}
To get these identities, we used
\begin{eqnarray}
&&
\sla v_H \sla v_H 
= 1
\, , \quad
\sla v_H \gam^\mu \sla v_H 
= 2 v_H^\mu \sla v_H- \gam^\mu
\, .
\end{eqnarray}

Assuming that the on-shell momentum $  m_H v_H^\mu$ 
does not change in the low-energy dynamics, we factorize 
the corresponding plane wave as 
\begin{eqnarray}
\Psi_\pm (x) \equiv e^{i m_H v_H^\mu x_\mu} \Q_\pm \Psi(x)
\, .
\end{eqnarray}
Then, we have 
\begin{eqnarray}
\Psi = e^{- i m_H v_H^\mu x_\mu}  \left( \, \Psi_+  + \Psi_-  \, \right ) 
\, .
\end{eqnarray}
Therefore, by using the identities (\ref{eq:Qid}), 
the kinetic term of the heavy quark is decomposed as
\begin{eqnarray}
\Lag = \bar \Psi  (i\sla D -m_H) \Psi 
\simeq
 \bar \Psi_+  \left( \,   i v_H^\mu D_\mu  \, \right) \Psi_+ 
\, .
\end{eqnarray}
At the leading order, the interaction vertex does not contain the gamma matrix, 
because the magnetic moment is suppressed by $ 1/m_H$. 
The free HQ propagator is read off as 
\begin{eqnarray}
S_H (k; \bv_H) = \frac{i}{v_H^\mu k_\mu + i \vep} \Q_+
\label{eq:Qprop}
\, .
\end{eqnarray}

\bibliography{Kondo_2SC_bib}

\end{document}